\documentclass[
reprint,
superscriptaddress,
nobibnotes,
amsmath,amssymb,
aps,
pra,
nofootinbib 
]{revtex4-1}

\usepackage{mathtools}
\usepackage{amsmath}
\usepackage{amsthm}
\usepackage{amssymb}
\usepackage{bm}
\usepackage{natbib}

\usepackage{xurl}
\usepackage{hyperref}
\hypersetup{breaklinks=true}

\newtheorem{defn}{Definition}

\newcommand{\C}{{\bf C}} 
\newcommand{\Smap}{{\bf S}} 
\newcommand{\A}{{\bf A}} 

\begin{document}

\title{On Principles and Representations for Extended Contextuality}

\author{Matt Jones (mcj@colorado.edu), University of Colorado}
\thanks{Thanks to Alisson Tezzin, Barbara Amaral, and Elie Wolfe for helpful comments.}
 
\begin{abstract}
There has been recent interest in whether the concept of quantum contextuality can be extended to systems with disturbance or signaling while retaining the essential properties of standard contextuality. 
\textcite{dzhafarov2023neither} (arXiv:2302.11995) offer a provocative argument that the answer is always affirmative:
If standard contextuality satisfies some principle that extended contextuality does not, then that principle must be `non-substantive' in that it depends on a superficial choice of representation.
Their argument is based on a \emph{consistification} procedure that maps all systems to nondisturbing ones and that preserves contextuality.
This paper raises several objections to their argument, including 
that it neglects how substantive principles change their expression under a change of representation,
that it begs the question by presuming the principles are based on non-substantive transformations, 
and that the argument applies universally to arbitrary extensions of contextuality.
\end{abstract}

\maketitle

Contexuality refers to a property of physical measurement systems in which certain observables can be measured in subsets (called contexts) but never all at once.
A system is contextual if the algebraic or probabilistic relations among observables within individual contexts are incompatible with a single model spanning all contexts \citep{kochen-specker-67,Abramsky_2011}.
This is equivalent to the condition that there exists no joint probability distribution over all the observables whose marginals match the distributions in individual contexts \citep{Fine1982}.
For example, if four binary observables can be measured only in pairs $\{q_1,q_2\}$, $\{q_2,q_3\}$, $\{q_3,q_4\}$, $\{q_1,q_4\}$, and all eight single-variable marginals are uniform ($\Pr[q_i=-1]=\Pr[q_i=1]=1/2$ for each $q_i$ in both contexts in which it is measured) while the bivariate joints satisfy $\Pr[q_1=q_2]=\Pr[q_2=q_3]=\Pr[q_3=q_4]=\Pr[q_1\ne q_4]=1$, then the four bivariate distributions are not compatible with any global four-way distribution \citep{popescu-rohrlich-94}.
Contextuality is significant because (under certain auxiliary assumptions) contextual systems are incompatible with the classical realist ontology whereby observables have values independent of being measured \citep{epr},
and because both the predictions of quantum mechanics \citep{bell64} and real quantum systems \citep{aspect1981experimental,hensen2015loophole} are contextual.

The concept of contextuality applies only to systems satisfying the condition of \emph{no disturbance} (or no signaling), which states that if some subset of observables $\boldsymbol{q}$ is jointly measured in distinct contexts $c$ and $c'$ then the distribution of $\boldsymbol{q}$ must be the same in both contexts. 
If this condition is violated then there is manifestly some causal connection between contexts and measurements, so there is nothing non-classical about the fact that the contexts are incompatible with a single distribution (e.g., if the value of $q_1$ depends on whether the experimenter also measures $q_2$ or $q_3$ then it is unsurprising that the two contexts $\{q_1,q_2\}$ and $\{q_1,q_3\}$ are incompatible).
In other words, contextuality is about incompatibility of contexts that is not explainable by (or that goes beyond) causal influence \citep{cervantes2018snow,bell64}.
Nevertheless, no-disturbance is difficult to achieve in real experiments, should not be expected in finite datasets because of sampling error, and is simply untrue of most physical systems outside particle physics.
Therefore there have been efforts to develop extended definitions of contextuality.
These extensions agree with standard contextuality---hereafter Kochen-Specker (KS) contextuality---on systems without disturbance and also apply to systems with disturbance \citep{PRL-Dzhafarov,DzhafarovCBD2.0}.

An important question for any extension of contextuality is whether it maintains essential properties of KS contextuality \citep{kujala2021contextuality,tezzin-impossibility-arxiv}.
For example, KS contextuality satisfies a \emph{monotonicity} property whereby measuring less information from a noncontextual system (e.g., by omitting an observable from a context or by coarse-graining its recorded values) cannot yield a contextual system. 
Contrapositively, measuring more information from a contextual system cannot yield a noncontextual system. 
This property is essential to the meaning of contextuality because of the concept's roots in questions of the completeness of quantum mechanics and more specifically of whether observed correlations can be explained by unobserved information \citep{epr,bell64}.
Therefore it is desirable that any extension of contextuality also satisfy monotonicity.

\textcite{dzhafarov2023neither} offer a provocative argument that claims to render these considerations moot.
Specifically, they introduce a broad class of extensions of contextuality named $\mathsf{C}$ contextuality and claim that any version of $\mathsf{C}$ contextuality (i.e., for any choice of $\mathsf{C}$) will satisfy any property that KS contextuality does, or else that property is automatically `non-substantive'.
The argument rests on a mapping $\C$ between measurement systems $P$ (which for precision we will call behaviors) 
that builds upon work by \textcite{amaral2018necessary} and satisfies the following properties:
(1) $\C$ is injective,
(2) $\C(P)$ is always nondisturbing even if $P$ is disturbing, and
(3) $\C(P)$ is KS-contextual if and only if $P$ is $\mathsf{C}$-contextual.
They claim that $\C$ is just a superficial change of representation called \emph{consistification} (for consistent connectedness, a term similar to no-disturbance). They further argue that under this mapping $\mathsf{C}$ contextuality is equivalent to a ``consistified'' theory that applies only to nondisturbing behaviors and that is identical to (or a restriction of) KS contextuality.
Therefore if $\mathsf{C}$ contextuality violates some proposed principle $\mathcal{A}$ (such as monotonicity) then either KS contextuality also violates $\mathcal{A}$ or else $\mathcal{A}$ is nonsubstantive because whether it is satisfied depends on a superficial choice of representation ($P$ vs.\ $\C(P)$).

This paper presents several counterarguments to \textcite{dzhafarov2023neither}.
First, the consistification mapping $\C$ is contrived and is a mere reformulation only in the sense that it is invertible (i.e., information-preserving).
Second, consistified $\mathsf{C}$ contextuality and KS contextuality are not actually the same theory, because while they agree in how they label certain mathematical systems as (non)contextual they disagree in how those mathematical systems represent physical systems.
Third, any principle $\mathcal{A}$ of interest will generally have a very different meaning when applied to the standard and consistified representations, so there is no reason to expect it to hold in the latter.
Fourth, claiming $\C$ is a mere reformulation requires assuming any properties it does not preserve are nonsubstantive, which begs the question.
Fifth, the consistification argument is nearly universally applicable, so if it is valid then it supports arbitrary, unprincipled extensions of contextuality just as well as it supports the principled ones that have been developed in the literature.
After presenting these arguments, we offer an analogy to elementary number theory to illustrate the plainly false implications of the consistification argument.

\section{Preliminaries}

We follow the standard model-independent formulation of contextuality \citep{Abramsky_2011,acin2015combinatorial,CSW, amaral2018graph,PRL-Dzhafarov}. A \emph{measurement scenario} is defined by a set of observables and the contexts in which they are measured.

\begin{defn}[Measurement scenario]\label{def:scenario}
    A measurement scenario is a tuple $\mathcal{S}=(\mathcal{Q},\mathcal{C},\prec,\mathcal{O})$ where $\mathcal{Q}$ is a set of observables, $\mathcal{C}$ is a set of contexts, $\prec\,\subset\!\mathcal{Q}\times\mathcal{C}$ is a measurement relation where $q\prec c$ indicates $q$ is measured in $c$, and $\mathcal{O}=(O_q)_{q\in\mathcal{Q}}$ where $O_q$ is the set of possible outcomes for $q$.
\end{defn}
We will implicitly assume $\mathcal{Q}$, $\mathcal{C}$, and $O_q$ are finite although the treatment can be readily extended to the general case.
We write $\boldsymbol{c}=\{q:q\prec c\}$ for the set of all observables measured in context $c$,
use boldface $\boldsymbol{q}$ to denote any subset of observables,
and write $\boldsymbol{q}\prec c$ as shorthand for $\forall_{q\in\boldsymbol{q}}\,q\prec c$.

A \emph{behavior} on $\mathcal{S}$ is a collection of joint outcome distributions for all contexts.

\begin{defn}[Behavior]\label{def:behavior}
    A behavior $P$ on a measurement scenario $\mathcal{S}$ is defined by a probability distribution $P(\cdot|c)$ for each $c\in\mathcal{C}$. 
    $P_{\boldsymbol{q}}(\boldsymbol{u}|c)$ indicates the probability that the observables in $\boldsymbol{q}$ have the joint outcome $\boldsymbol{u}$ for any $\boldsymbol{q}\prec c$ and $\boldsymbol{u}\in\prod_{q\in\boldsymbol{q}}O_q$.
\end{defn}

A behavior is nondisturbing if the distributions for distinct contexts agree on overlapping subsets of observables.

\begin{defn}[Disturbance]
    A behavior $P$ is nondisturbing if for every $c,c'\in\mathcal{C}$ and $\boldsymbol{q}\subset\boldsymbol{c}\cap\boldsymbol{c}'$ it satisfies $P_{\boldsymbol{q}}(\cdot|c)=P_{\boldsymbol{q}}(\cdot|c')$. Otherwise $P$ is disturbing.
\end{defn}

A behavior is KS-noncontextual if there exists a global distribution over all of $\mathcal{Q}$ whose marginals match the observed distributions in all contexts. This definition, due to \textcite{Fine1982}, is equivalent to the original \textcite{kochen-specker-67} definition in terms of hidden-variable models.

\begin{defn} [KS contextuality] \label{def:KS-contextuality}
    A behavior $P$ is KS-noncontextual if there exists a distribution $\bar{P}$ on $\prod_{q\in\mathcal{Q}}O_q$ such that $\bar{P}_{\boldsymbol{c}}(\cdot)=P(\cdot|c)$ for all $c\in\mathcal{C}$. Otherwise $P$ is KS-contextual.
\end{defn}

To facilitate the analysis below, we introduce the indicator function $\mathsf{T}_{\rm KS}$ where $\mathsf{T}_{\rm KS}(P)=0$ if $P$ is KS-noncontextual and $\mathsf{T}_{\rm KS}(P)=1$ if $P$ is KS-contextual. The domain of $\mathsf{T}_{\rm KS}$ is the class of all nondisturbing behaviors $P$. An \emph{extension of contextuality} is any function of this form with domain including disturbing behaviors.

\begin{defn}[Extension of contextuality]
    An extension of contextuality is a function $\mathsf{T}:\mathfrak{R}\to\{0,1\}$ where $\mathfrak{R}$ is some class of disturbing and non-disturbing behaviors,
    such that $\mathsf{T}(P)=\mathsf{T}_{\rm KS}(P)$ for all nondisturbing $P\in\mathfrak{R}$.
    We say $P$ is $\mathsf{T}$-contextual if $\mathsf{T}(P)=1$ and $\mathsf{T}$-noncontextual if $\mathsf{T}(P)=0$.
\end{defn}

For a complete extension of contextuality, $\mathfrak{R}$ would be the class of all behaviors, but we allow it to be any subclass (e.g., ones with only binary observables \citep{DzhafarovCBD2.0}).

We next describe three properties that KS contextuality satisfies \citep{tezzin-impossibility-arxiv} and are arguably desirable properties of any extension of contextuality. The first two were introduced in \cite{kujala2021contextuality} and are forms of monotonicity, which is described in the introduction and argued there to be essential to the standard meaning of contextuality.

\begin{defn}[Nestedness]
    Measurement scenarios $\mathcal{S}$ and $\mathcal{S}'$ are nested if $\mathcal{Q}'\subset\mathcal{Q}$, $\mathcal{C}'\subset\mathcal{C}$, and $\prec'\,\subset\,\prec$, meaning $q\prec c$ whenever $q\prec'c$. 
    Behaviors $P$ on $\mathcal{S}$ and $P'$ on $\mathcal{S}'$ are nested if $P_{\boldsymbol{q}}(\cdot|c)=P'_{\boldsymbol{q}}(\cdot|c)$ whenever $\boldsymbol{q}\prec'c$ (i.e., $P'$ is defined by the appropriate marginals of $P$).
    An extension of contextuality $\mathsf{T}$ satisfies Nestedness if $\mathsf{T}(P')\le\mathsf{T}(P)$ whenever $P'$ is nested in $P$ (i.e., if $P$ is $\mathsf{T}$-noncontextual then so is $P'$).
\end{defn}

\begin{defn}[Coarse-graining]
    Scenario $\mathcal{S}'$ is a coarse-graining of $\mathcal{S}$ if $\mathcal{Q}'=\mathcal{Q}$, $\mathcal{C}'=\mathcal{C}$, $\prec'=\prec$, and for every $q\in\mathcal{Q}$, $O'_q=g_q(O_q)$ for some function $g_q$. Thus $\mathcal{S}'$ may fail to distinguish some outcomes in $\mathcal{S}$ (i.e., when $g_q(u_1)=g_q(u_2)$). 
    Behavior $P'$ on $\mathcal{S}'$ is a coarse-graining of $P$ on $\mathcal{S}$ if $P'(\boldsymbol{u}|c)=P(\boldsymbol{g}_{\boldsymbol{c}}^{-1}(\boldsymbol{u})|c)$ for all $c$ (i.e., $P'(\cdot|c)$ is the push-forward of $P(\cdot|c)$ by the composite function $\boldsymbol{g}_{\boldsymbol{c}}$).
    An extension of contextuality $\mathsf{T}$ satisfies Coarse-graining if $\mathsf{T}(P')\le\mathsf{T}(P)$ whenever $P'$ is a coarse-graining of $P$ (i.e., if $P$ is $\mathsf{T}$-noncontextual then so is $P'$).
\end{defn}

The third property we consider is based on post-processing, in which is a new observable is defined by a (classical, deterministic) function of existing observables. KS contextuality is closed under post-processing in the sense that post-processing cannot create a KS-contextual behavior from a KS-noncontextual one. It is closely related to Kochen and Specker's original question of whether observables can be written as deterministic functions of some hidden variable, based on the functional relationships among them \citep{kochen-specker-67}. Likewise, resource theories of contextuality define post-processing as a free operation that cannot increase the degree of contextuality \cite{shane2019comonadic,Karvonen2019categories}.

\begin{defn}[Post-processing]
    A post-processing of a set of observables $\boldsymbol{q}=\{q_1,\dots,q_n\}$ is a function $f(q_1,\dots,q_n)$ with domain $\prod_{i=1}^n O_{q_i}$. 
    This post-processing defines an expanded scenario $\mathcal{S}'$ with $\mathcal{Q}'=\mathcal{Q}\cup\{q'\}$ and $q'\prec' c$ iff $\boldsymbol{q}\prec c$ for all $c$. 
    The post-processing of a behavior $P$ is a behavior $P'$ on $\mathcal{S}'$ satisfying $P'(q'=f(q_1,\dots,q_n)|c)=1$ for all $c\succ\boldsymbol{q}$. That is, $P'_{\boldsymbol{q},q'}(\boldsymbol{u},u|c)=1$ iff $u=f(\boldsymbol{u})$. 
    An extension of contextuality $\mathsf{T}$ satisfies Post-processing if $\mathsf{T}(P')\le\mathsf{T}(P)$ whenever $P'$ is a post-processing of $P$ (i.e., if $P$ is $\mathsf{T}$-noncontextual then so is $P'$).
\end{defn}

Nestedness, Coarse-graining, and Post-processing are all consistency properties whereby certain transformations cannot produce contextual behaviors from non-contextual ones. Because these properties are all central to the concept of KS contextuality, it seems of interest to know whether there exist extensions of contextuality that also satisfy them.

\section{The consistification argument}\label{sec:consistification}

\textcite{dzhafarov2023neither} attempt to nullify any discussion of whether extensions of contextuality can satisfy properties of KS contextuality, through the following construction.
The construction comprises a family of extensions of contextuality and, for each extension, an injective map from the domain of that extension ($\mathfrak{R}$) to the subclass of non-disturbing behaviors.

Let $\mathsf{C}$ be some criterion on joint probability distributions that identifies a unique joint distribution given its marginals. More precisely, assume $\mathsf{C}$ has the following Uniqueness property:

\begin{defn}[Uniqueness criterion]\label{def:uniqueness}
    Assume we are given $n$ copies of an observable $q$ and specifications of the marginal distributions $P_{q_i}(\cdot)$ for all $i=1,\dots,n$, and that these marginals collectively are compatible with $\mathfrak{R}$.
    Then a criterion $\mathsf{C}$ on joint distributions on $O_q^n$ has the Uniqueness property if
    \begin{enumerate}
        \item there is a unique distribution $P$ on $O_q^n$ having the given marginals that satisfies $\mathsf{C}$, and \label{crit:unique-joint}
        \item if the marginals are all the same, $P_{q_1}(\cdot)=\dots=P_{q_n}(\cdot)$, then the unique distribution satisfying $\mathsf{C}$ is the one in which all variables are equal, $P(q_1=\dots=q_n)=1$.\label{crit:all-equal}
    \end{enumerate}
\end{defn}

Under these assumptions, \textcite{dzhafarov2023neither} define an extension of contextuality named $\mathsf{C}$ contextuality.

\begin{defn}[$\mathsf{C}$ contextuality]\label{def:C-contextuality}
    Provided $\mathsf{C}$ has the Uniqueness property, $P\in\mathfrak{R}$ is $\mathsf{C}$-noncontextual iff there exists a distribution $\bar{P}$ on $\prod_{(q,c):q\prec c}O_q$ such that
    \begin{enumerate}
        \item \label{crit:match-bunches} $\bar{P}$ matches $P$ within each context: for each $c$, $\bar{P}_{\{(q,c):q\prec c\}}(\cdot)=P(\cdot|c)$ under the natural correspondence of indices $(q,c)\leftrightarrow q$
        \item \label{crit:match-connections} for each $q$ the joint distribution 
        $\bar{P}_{\{(q,c):c\succ q\}}(\cdot)$ satisfies $\mathsf{C}$.
    \end{enumerate}

\end{defn}

To be clear, $\bar{P}$ is a distribution not over the original observables $q\in\mathcal{Q}$ (as in Definition \ref{def:KS-contextuality}) but over a new set of observables indexed by the pairs $(q,c)$ where each such pair corresponds to a copy of (i.e., has the same outcome set as) $q$. If we imagine an incomplete $\mathcal{C}\times\mathcal{Q}$ matrix containing only the cells satisfying $q\prec c$ then the marginal of $\bar{P}$ corresponding to each row $c$ is fully determined by $P(\cdot|c)$ (Definition \ref{def:C-contextuality}.\ref{crit:match-bunches}) and the marginal of $\bar{P}$ corresponding to each column $q$ is fully determined by $\mathsf{C}$ (Definition \ref{def:C-contextuality}.\ref{crit:match-connections}) due to $\mathsf{C}$'s Uniqueness property. Thus $\mathsf{C}$ contextuality corresponds to the question of whether these row and column distributions can be combined into one master distribution.

Note that Definition \ref{def:uniqueness}.\ref{crit:all-equal} implies $\mathsf{C}$ contextuality must agree with KS contextuality on nondisturbing behaviors. This is because any $\bar{P}$ satisfying Definition \ref{def:C-contextuality} for a nondisturbing $P$ can be converted to a global distribution on $\mathcal{Q}$ in the sense of Definition \ref{def:KS-contextuality} by collapsing the column $\{(q,c):c\succ q\}$ for each $q$ into a single variable \cite{PRL-Dzhafarov}.

As an example, let $\mathfrak{R}$ be the class of behaviors in which all observables are binary, and let $\mathsf{C}$ be the property that for all $i,j$, $P(q_i=q_j)$ takes the maximum possible value given $P_{q_i}(\cdot)$ and $P_{q_j}(\cdot)$. Then $\mathsf{C}$-contextuality corresponds to CBD 2.0 \cite{DzhafarovCBD2.0}.
    
The second part of the construction in Ref.\ \cite{dzhafarov2023neither} draws on the insight that $\bar{P}$ in Definition \ref{def:C-contextuality} can be viewed as a global distribution in the sense of Definition \ref{def:KS-contextuality}, where the ``observables'' are the pairs $(q,c)$ with $q\prec c$ and the ``contexts'' correspond to the row and column distributions described above.
This is formalized as follows.

\begin{defn}[Consistification]\label{def:consistification}
    Given a secenario $\mathcal{S}$, the consistified scenario $\tilde{\mathcal{S}}$ is defined by $\tilde{\mathcal{Q}}=\{(q,c)\in\mathcal{Q}\times\mathcal{C}:q\prec c\}$ and $\tilde{\mathcal{C}}=\mathcal{Q}\cup\mathcal{C}$. 
    We write $\Tilde{q}$ and $\Tilde{c}$ as the elements of $\tilde{\mathcal{C}}$ corresponding to $q$ and $c$. 
    Every observable in $\tilde{\mathcal{S}}$ belongs to exactly two contexts: $(q,c)\,\tilde{\prec}\,\tilde{q}$ and $(q,c)\,\tilde{\prec}\,\tilde{c}$. 

    Given a behavior $P$ on $\mathcal{S}$, the consistification $\Tilde{P}$ is a behavior on $\tilde{\mathcal{S}}$ defined as follows. For every $c\in\mathcal{C}$, $\Tilde{P}(\cdot|\Tilde{c})=P(\cdot|c)$ via the bijection of indices $(q,c)\leftrightarrow q$ for $q\prec c$. That is, $\Tilde{P}_{(\boldsymbol{q},c)}(\cdot|\Tilde{c})=P_{\boldsymbol{q}}(\cdot|c)$ for all $\boldsymbol{q}\prec c$, where $(\boldsymbol{q},c)=\{(q,c):q\in\boldsymbol{q}\}$. 
    For every $q\in\mathcal{Q}$, $\tilde{P}(\cdot|\tilde{q})$ is the unique distribution satisfying $\mathsf{C}$ with marginals matching $P$: $\tilde{P}_{(q,c)}(\cdot|\tilde{q})=P_q(\cdot|c)$ for all $c\succ q$. 
\end{defn}

Thus the context $\Tilde{c}$ contains all observables $(q,c)$ for which $q\prec c$ (ranging over $q$), and the context $\Tilde{q}$ contains all observables $(q,c)$ for which $q\prec c$ (ranging over $c$).
The distributions $\tilde{P}(\cdot|\tilde{c})$ and $\tilde{P}(\cdot|\tilde{q})$ are respectively determined by criteria \ref{crit:match-bunches} and \ref{crit:match-connections} in Definition \ref{def:C-contextuality}.

For convenience we define the consistification mapping $\C$, from behaviors to behaviors, by $\C(P)=\tilde{P}$. \textcite{dzhafarov2023neither} show that $\C$ has the following properties:
(1) $\C$ is injective, 
(2) $\C(P)$ is always nondisturbing, and
(3) $\C(P)$ is KS-contextual iff $P$ is $\mathsf{C}$-contextual.
Based on property 1, they argue $\C$ is a ``mere reformulation'', meaning $P$ and $\C(P)$ are two ways of mathematically representing the same information, and the choice is arbitrary and superficial. Based on properties 2 and 3, they argue $\mathsf{C}$ contextuality is equivalent to (a subtheory of) KS contextuality, because the statements $P$ is $\mathsf{C}$-contextual and $\C(P)$ is KS-contextual are logically equivalent, merely expressed in terms of different representations. 
As a corollary, they argue that any property KS contextuality satisfies and $\mathsf{C}$ contextuality violates must be non-substantive, because whether the property holds is just a matter of one's choice of representation. Therefore there is no reason to ask whether an extension of contextuality satisfies a property such as Nestedness, Coarse-graining or Post-processing. Either it does and there is no problem, or else it does not but then this implies the property is non-substantive and should not be of interest in the first place.

\section{Counterarguments}

\subsection{Properties of the consistified representation} \label{sec:contrived}

The consistified representation is very different from the standard one. 
It includes multiple copies of each original observable (one for each original context it appears in) and introduces pseudo-contexts ($\tilde{q}$) to represent joint distributions that were not actually observed.
Discussion of how natural some representation is is inherently subjective, but it seems clear that the only sense in which the consistified representation can be considered a mere reformulation of the standard one is that they contain the same information; that is, that $\C$ is injective and hence invertible. This will be relevant to our universality argument below.

\subsection{Mathematical vs.\ physical theories}\label{sec:math-phys}

Let $\mathsf{T_C}:\mathfrak{R}\to\{0,1\}$ be the indicator function for $\mathsf{C}$ contextuality, paralleling the definition above of $\mathsf{T}_{\rm KS}$. 
Define $\mathsf{T}'_{\mathsf{C}}:\C(\mathfrak{R})\to\{0,1\}$ by $\mathsf{T}'_{\mathsf{C}}(\C(P))=\mathsf{T_C}(P)$.
This is the indicator function for consistified $\mathsf{C}$ contextuality.
The consistification argument holds that $\mathsf{T_C}$ and $\mathsf{T}'_{\mathsf{C}}$ embody essentially the same theory, differing only in a choice of how behaviors are represented. 
Property 1 of $\C$ (injectivity) implies $\mathsf{T}'_{\mathsf{C}}$ is well-defined, and property 2 implies its domain contains only nondisturbing behaviors.
Property 3 of $\C$ implies $\mathsf{T}'_{\mathsf{C}}$ is the restriction of $\mathsf{T}_{\rm KS}$ to $\C(\mathfrak{R})$.
Therefore, the argument goes, $\mathsf{C}$ contextuality ($\mathsf{T_C}$) is equivalent to a subtheory of KS contextuality ($\mathsf{T}_{\rm KS}$).

Our first counterargument is that consistified $\mathsf{C}$ contextuality and KS contextuality are not actually the same theory.
To be rigorous about the notion of change of representation, we must be explicit about what is being represented. Let $\mathfrak{B}$ be the class of physical behaviors or phenomena under study. The elements of $\mathfrak{B}$ are not mathematical objects but real or imagined experiments, or perhaps sets of random variables in a more abstract setting. To say these are represented by the mathematical objects we call measurement scenarios and behaviors is to posit a map from $\mathfrak{B}$ to $\mathfrak{R}$. Let $\Smap:\mathfrak{B}\to\mathfrak{R}$ be the standard representation implicit in Definitions \ref{def:scenario} and \ref{def:behavior}, so that $\C\circ\Smap$ is the consistified one. 
Then 
$\mathsf{C}$ contextuality is embodied by the map $\mathsf{T_C}\circ\Smap:\mathfrak{B}\to\{0,1\}$, 
consistified $\mathsf{C}$ contextuality is embodied by $\mathsf{T}'_\mathsf{C}\circ\C\circ\Smap$, and 
KS contextuality is embodied by $\mathsf{T}_{\rm KS}\circ\Smap$.
Therefore $\mathsf{C}$ contextuality and consistified $\mathsf{C}$ contextuality are indeed the same theory ($\mathsf{T_C}\circ\Smap=\mathsf{T}'_\mathsf{C}\circ\C\circ\Smap$) but KS contextuality is different. 
KS contextuality and consistified $\mathsf{C}$ contextuality agree in how they classify systems of probability distributions as contextual or noncontextual ($\mathsf{T}'_{\mathsf{C}}=\mathsf{T}_{\rm KS}\vert_{_{\C(\mathfrak{R)}}}$), yet they differ in how they use systems of probability distributions to represent systems of physical measurements ($\Smap$ vs.\ $\C\circ\Smap$).

The consistification argument holds that no substantive property can hold for one theory and not another when those theories differ only in a choice of representation. The preceding analysis shows that this applies to the comparison between $\mathsf{C}$ contextuality and consistified $\mathsf{C}$ contextuality, but not to a comparison between either of these and KS contextuality. In particular, KS contextuality and consistified $\mathsf{C}$ contextuality apply the same superificial criterion (Definition \ref{def:KS-contextuality}) to the same mathematical objects (nondisturbing $P$), but those objects have different meanings in the two theories. Therefore there may be substantive properties satisfied by one theory and not the other.



\subsection{Distortion of principles}\label{sec:distort}

The principles we consider all take the form that if behavior $P$ is noncontextual then so is behavior $\A(P)$, where $\A$ belongs to a class of transformations defined for example by Nestedness, Coarse-graining, or Post-processing.  
A notable consequence of the disparity between KS contextuality and consistified $\mathsf{C}$ contextuality is that these transformations manifest in peculiar ways in the latter theory. For instance, while post-processing under the standard representation ($\Smap$) involves appending a new observable (as is sensible), the consistified representation ($\C\circ\Smap$) is compelled to append new pseudo-contexts. 

This observation reflects the fact that consistification does not respect the physical structure of measurement systems. For instance, consistifying a coarse-graining of a behavior does not yield a coarse-graining of its consistification.
Mathematically, $\A$ and $\C$ do not commute.
A transformation $\A$ applied to the standard representation corresponds to $\C\A\C^{-1}$ in the consistified representation.
This implies that consistified $\mathsf{C}$ contextuality may violate some principle that KS contextuality satisfies, and consequently that $\mathsf{C}$ contextuality may satisfy that principle for nondisturbing behaviors but violate it for disturbing ones.

To explain this in more detail: KS contextuality is known to satisfy the principles in question (Nestedness, Coarse-graining, Post-processing, and others) \cite{tezzin-impossibility-arxiv}.
Therefore $\A$ preserves KS noncontextuality and also $\mathsf{C}$ noncontextuality when applied to nondisturbing behaviors (where the two theories agree).
However, $\C\A\C^{-1}$ does not have to: there may exist a nondisturbing $\tilde{P}$ that is KS-noncontextual (hence also $\mathsf{C}$-noncontextual) while $\C\A\C^{-1}(\tilde{P})$ is KS-contextual (and $\mathsf{C}$-contextual).%
\footnote{Indeed, Theorem 2 of \citet{tezzin-impossibility-arxiv} implies there must exist such a $\tilde{P}$ for every choice of $\mathsf{C}$, for at least one of the transformations $\A$ they consider.}
Write $P=\C^{-1}(\tilde{P})$.
Because $\C$ and $\C^{-1}$ preserve $\mathsf{C}$ contextuality, $P$ is $\mathsf{C}$-noncontextual while $\A(P)$ is C-contextual.
This also implies $P$ is disturbing (since $\A$ cannot create contextuality for nondisturbing behaviors).
Therefore $\mathsf{C}$ contextuality violates the principle embodied by $\A$ for some disturbing behaviors.

In summary, once we recognize that the change of representation changes the mathematical realization of a substantive principle or transformation ($\A$ to $\C\A\C^{-1}$), the argument that $\mathsf{C}$ contextuality and KS contextuality must obey the same substantive principles breaks down.

\subsection{Begging the question}\label{sec:beg}

Anticipating the argument in the preceding subsection, \citet{dzhafarov2023neither} assert that $\C$ should not be expected to preserve the transformations in question. 
For example, consistifying a coarse-graining of a behavior should not be expected to yield a coarse-graining of its consistification.
However, if $\C$ is itself considered  substance-preserving—a `mere reformulation'  of the original system—while not respecting these other transformations of nesting, coarsening, post-processing, etc., then one must take the  debatable position that the latter transformations are not substantive or physically meaningful. Moreover, the nonsubstantiveness of these transformations (or the principles based on them) is precisely what the consistification argument aims to establish, making it circular. 

Another way to articulate our argument is that the consistency properties in question are  intended to apply to physical behaviors ($B\in\mathfrak{B}$), not their mathematical representations ($P\in\mathfrak{R}$). This  stance aligns with the position taken in Ref.\ \cite{dzhafarov2023neither}. If $P_1$ represents $B_1$ under the standard representation, and $\tilde{P}_1$ represents $B_1$ under the consistified one, and we then apply some physically meaningful transformation to obtain $P_2$ and $\tilde{P}_2$, one should expect $P_2$ and $\tilde{P}_2$ to represent the same physical system $B_2$. (This  expectation is equivalent to the requirement that $\A$ and $\C$ commute for any physically meaningful $\A$.)
This is not the case with consistification and the transformations we consider.
Therefore, the assertion that consistification is a mere reformulation requires positing that the transformations—nesting, coarse-graining, post-processing, etc.—are not physically meaningful. 
Moreover, this claim must be made as a prerequisite to the consistification argument, to justify the use of $\C$ in the first place.
Because there is little difference between the premise that $\A$ is not physically meaningful and the conclusion that the principle based on $\A$ (i.e., that it cannot create contextuality) is not substantive, the entire consistification argument becomes circular.

\subsection{Universal applicability}\label{sec:universal}

Finally,  by discarding any requirement for $\C$ to respect the physical properties of measurement systems or their representations, the consistification argument  becomes universally applicable. The consistification procedure embodies no principles other than the extension of contextuality it begins with (see the discussion immediately preceding Definition \ref{def:consistification}, linking it to Definition \ref{def:C-contextuality}).  While it requires $\C$ to be a `mere reformulation', 
 no specific definition of this criterion is offered other than that $\C$ is injective (see Section \ref{sec:contrived}).

Given an arbitrary extension of contextuality $\mathsf{T}$, one could always define a mapping $\C$ that satisfies the three properties given  in Section \ref{sec:consistification} (injective, nondisturbing output, and contextuality preserving), since the sets of disturbing, nondisturbing KS-contextual, and nondisturbing KS-noncontextual behaviors on finite measurement scenarios (up to relabeling of $\mathcal{Q,C,O}$) all have the same infinite cardinality. This can be done for any extension, not just ones fitting Definition \ref{def:C-contextuality}. Then following the consistification argument, any basis for rejecting $\mathsf{T}$ must either also reject KS contextuality or else be `non-substantive' because it depends on whether one chooses the standard representation of behaviors $\Smap$ or the representation $\C\circ\Smap$. This is a far more damaging conclusion for the field of extended contextuality than the one \citet{dzhafarov2023neither} aim to avoid, which is that no extension can simultaneously satisfy several core properties of KS contextuality \cite{tezzin-impossibility-arxiv}. Instead the conclusion from the consistification argument would be that all extensions are valid, even absurd ones, and there can be no substantive basis for choosing among them.

\section{Number-theoretic illustration}



To illustrate how the logic of the consistification argument would apply in another example,
suppose Alice proposes a theory $\mathsf{T}$ of natural numbers where $\mathsf{T}$ is the statement that $n$ is even iff $n$ is 2 or nonprime. Alice  argues in support of $\mathsf{T}$ as follows. She defines $\mathbb{M} \subset \mathbb{N}$ as the set of all primes and multiples of 2, and $\mathsf{T}^{\prime}$ as the restriction of $\mathsf{T}$ to $\mathbb{M}$, and observes that $\mathsf{T}^{\prime}$ is universally regarded as correct (i.e., it agrees with the standard definition of even numbers). She then defines the injective function $\C:\mathbb{N}\to\mathbb{M}$ by $\C\left(n\right)=2^{\sigma\left(n\right)-2}n$ where $\sigma\left(n\right)$ is the number of divisors of $n$ (with the exception $\C(1)=1$), and observes that any $n\in\mathbb{N}$ is even according to $\mathsf{T}$ iff $\C\left(n\right)$ is even according to $\mathsf{T}^{\prime}$; that is, $\mathsf{T}=\mathsf{T}'\circ\C$. 
Because $\bf C$ is invertible, $n\leftrightarrow\C\left(n\right)$ is just a change of representation, making $\mathsf{T}$ and $\mathsf{T}'$ mere reformulations of each other.
Therefore, because $\mathsf{T}^{\prime}$ is universally accepted, $\mathsf{T}$  should be as well. 

If, Alice continues, one chooses to reject $\mathsf{T}$ but retain $\mathsf{T}^{\prime}$, the basis for that conclusion must not be substantive because it is applied under one representation ($n$) and not under another ($\C\left(n\right)$). 
For example, suppose Bob proposes the axiom $\mathcal{A}$: if $n$ is even then so is $\A(n)=n+2$. Bob notes that $\mathcal{A}$ contradicts $\mathsf{T}$ but not $\mathsf{T}^{\prime}$, thus offering a basis for rejecting the former while retaining the latter. Alice replies that, if $\mathcal{A}$ were substantive, then Bob should apply it identically under the two representations. 
Moreover, this observation would apply to any basis (not just $\mathcal{A}$) for rejecting $\mathsf{T}$ while retaining $\mathsf{T}'$, and so  there can be no substantive basis for accepting $\mathsf{T}'$ while not accepting $\mathsf{T}$.
Alice's reasoning is isomorphic to the consistification argument (even $\leftrightarrow$ noncontextual, $\mathsf{T}\leftrightarrow\mathsf{C}\text{ contextuality}$, $\mathsf{T^{\prime}}\leftrightarrow\text{consistified $\mathsf{C}$ contextuality}$, $\mathbb{M}\leftrightarrow\text{nondisturbing behaviors},$ $\C\leftrightarrow\text{consistification}$, $\mathcal{A}\leftrightarrow\text{Nestedness, Coarse-graining, Post-processing, etc.}$), and it shows how that argument can produce clearly false conclusions. Unlike the case of contextuality, there is an extension of evenness from $\mathbb{M}$ to $\mathbb{N}$ that is universally regarded as correct, and yet Alice's reasoning still leads her to endorse a different, objectively false extension.

One counter to Alice's argument, paralleling Section \ref{sec:math-phys}, is that she identifies $\mathsf{T}'$ as a restriction of $\mathsf{T}$ to $\mathbb{M}$ with $\mathsf{T}'$ as the pullback of $\mathsf{T}$ through $\C$, $\mathsf{T}\circ\C^{-1}$. These coincide mathematically, but they are not the same theory any more than KS contextuality and consistified $\mathsf{C}$ contextuality are.
Consequently, she mistakenly takes $\mathcal{A}$ to have the same meaning in both representations. Paralleling Section \ref{sec:distort}, accounting for the change in representation yields $\mathcal{A}^{\prime}$: if $\C\left(n\right)$ is even then so is $\C\left(n+2\right)$ (i.e., if $m$ is even then so is $\C\A\C^{-1}(m)$). Clearly $\mathcal{A}$ and $\mathcal{A}^{\prime}$ are different statements. In particular, $\mathcal{A}$ is true while $\mathcal{A}^{\prime}$ is not, because the operation of adding $2$ manifests  differently under the nonstandard representation (e.g., $\C\left(9\right)=18$ while $\C\left(9+2\right)=11$). 
If Alice replies that the operation of adding 2 should not be expected to be preserved under $\bf C$ (paralleling Section \ref{sec:beg}), then  this together with the claim that $\C$ is a mere reformulation entails an assertion that adding $2$ is not a natural or substantive operation.  Moreover, this is begging the question, because the eventual conclusion is that $\mathcal{A}$---which concerns consistency under addition of 2---is nonsubstantive.

This example, which mirrors the logical structure of the consistification argument \cite{dzhafarov2023neither}, illustrates that substantive principles will generally change their meaning under a change of representation. Therefore, whether a given theory is consistent with those principles can very well change if the change of representation does not respect (i.e., commute with) the relational structure they invoke.

\section{Conclusions}
 
The goal of extending theories of contextuality to disturbing systems is an important one. Valuable progress has been made, most notably in the development of Contextuality by Default (CbD) \cite{PRL-Dzhafarov,DzhafarovCBD2.0}. Nevertheless, CbD does not carry the no-go implications that KS contextuality and Bell nonlocality do \cite{Jones2019}. Therefore it is unclear what conclusions can be drawn about a physical system based on a determination of CbD contextuality. This situation is especially concerning because physicists have begun to use CbD to analyze experimental data or theoretical scenarios \cite{bacciagaluppi2014leggett,fluhmann2018sequential,malinowski2018probing,kupczynski2021contextuality,amaral2018necessary,wang2022significant,Andrei2022complementarity,zhan2017experimental}.

One promising way forward is to investigate whether extensions of contextuality satisfy the properties that are essential to KS contextuality \cite{kujala2021contextuality,tezzin-impossibility-arxiv}. Such efforts may guide development of more informative extensions and understanding of their implications for real systems. At first glance, the consistification argument might appear to make these considerations unnecessary, by showing that any version of $\mathsf{C}$ contextuality is just as valid as KS contextuality. However, even if the argument were valid it would not answer the question of what we learn about a disturbing system by determining it is $\mathsf{C}$-contextual (for any $\mathsf{C}$). On second consideration, taking into account the universal applicability explained in Section \ref{sec:universal}, the consistification argument might appear to undermine the entire enterprise, by showing that all extensions of contextuality (even absurd ones) are equally valid and there can be no basis for choosing among them. However, a more careful analysis of what it means for two theories to express the same principles in different representations (Sections \ref{sec:math-phys}-\ref{sec:beg}) reveals that the consistification argument simply does not prove what it sets out to prove. In particular, it is possible for theories that apply the same superficial criteria using different representations (as with KS contextuality and consistified $\mathsf{C}$ contextuality) to disagree on substantive principles. Therefore the consistification argument in no way jeopardizes the research program of identifying which substantive principles are satisfied by various extensions of contextuality.

\bibliographystyle{apsrev4-1}
\bibliography{ConsistifyReply.bib}

\end{document}